\documentstyle[sprocl]{article}

\input{psfig}

\begin{document}

\title{TEST OF GAUGE COVARIANCE OF FERMION-PHOTON VERTEX IN QUENCHED, MASSLESS THREE DIMENSIONAL QUANTUM ELECTRODYNAMICS}

\author{PAULUS C. TJIANG and CONRAD J. BURDEN}
\address{Department of Theoretical Physics, \\ Research School of Physical Sciences and Engineering, \\ The Australian National University, Canberra ACT 0200, Australia \\ E-mail : pct105@quantum.anu.edu.au, conrad.burden@anu.edu.au}

\maketitle

\abstracts{We study the gauge covariance of the fermion-photon vertex in quenched, massless 3-dimensional quantum electrodynamics. A class of vertex Ans\"{a}tze obtained by generalizing that proposed by Dong et al. is tested using the invariance of the photon polarization scalar under the Landau-Khalatnikov transformation.}

\section{Introduction}

Quantum electrodynamics in two spatial and one temporal dimensions (QED$_3$) is an ideal model for studying non-perturbative field theory. It has confinement, dynamical chiral symmetry breaking and is a superrenormalizable abelian theory.

One non-perturbative approach for studying QED$_3$ is to use the Schwinger-Dyson equations (SDEs). Since the full set of SDEs for any field theory contains an infinite tower of coupled non-linear integral equations, one generally makes a suitable Ansatz for the fermion-photon vertex. This reduces the full set to a finite set of coupled integral equations for the fermion and gauge-boson propagators. There have been many attempts to construct such a vertex Ansatz. One such attempt, by Dong et al.~\cite{DMR94}, uses a condition called the {\it transverse condition}~\cite{BR93} to restrict the Ball-Chiu vertex in quenched, massless QED.

In this paper we shall discuss the transverse condition for quenched, massless QED$_3$. We shall discuss briefly the fermion SDEs in Section \ref{fermionSDE}. The construction of a suitable fermion-photon vertex Ansatz will be discussed in Section \ref{vertex}. Use of vacuum polarization scalar for testing the gauge covariance of fermion-photon vertex is discussed in Section \ref{Photonscalar}, and conclusions are drawn in Section \ref{summary}.

\section{Schwinger-Dyson Equations for Quenched, Massless QED$_3$}  \label{fermionSDE}

In Euclidean space, the fermion SDE for quenched, massless QED$_3$ is

\begin{equation}
1 = i \not\! p S(p) + e^2 \int \frac{d^3q}{(2\pi)^3} D_{\mu \nu} (p - q) \gamma_\mu S(q) \Gamma_\nu (q,p) S(p),		\label{fermionSDE2}
\end{equation}
where the photon and fermion propagators are
\begin{equation}
D_{\mu \nu} (p,\xi) = \frac{1}{p^2} \left( \delta_{\mu \nu} - \frac{p_\mu p_\nu}{p^2} \right) + \xi \frac{p_\mu p_\nu}{p^4} = D_{\mu \nu}^T (p) +\xi  \frac{p_\mu p_\nu}{p^4},   \label{photon}
\end{equation}
and
\begin{equation}
S (p)^{-1} = i \not\!p A(p^2) + B(p^2),    \label{fermion}
\end{equation}
respectively. Lorentz covariance and the Ward-Takahashi identity restrict the fermion photon vertex $\Gamma_\mu (p,q)$ to be of the form
\begin{equation}
\Gamma_\mu (p,q) = \Gamma_\mu^{BC} (p,q) + \Gamma_\mu^T (p,q),  \label{fullvertex}
\end{equation}
given by Ball and Chiu~\cite{BC80}. In the chirally symmetric case ($B \equiv 0$) considered here
\begin{equation}
\Gamma_\mu^{BC} (p,q) = \frac{1}{2} \left[ A(p) + A(q) \right] \gamma_\mu + \frac{(p + q)_\mu}{p^2 - q^2} \left[ A(p) - A(q) \right] \frac{\not\!p + \not\!q}{2},   \label {BallChiu}
\end{equation}
and the transverse part of the vertex $\Gamma_\mu^T (p,q)$ satisfies
\begin{eqnarray}
(p - q)_\mu \Gamma_\mu^T (p,q) = 0 & \mbox{;} & \Gamma_\mu^T (p,p) = 0.  \label{transversevertex1}
\end{eqnarray}

\section{Gauge Covariant Fermion-Photon Vertex : Conditions and Constructions} 
\label{vertex}

It has traditionally been common practice in SDE studies of QED$_3$ or QED$_4$ to assume that in the quenched, massless limit, the fermion SDEs admit a trivial solution of bare fermion propagator and bare vertex in Landau gauge~\cite{DMR94,BR93,CP91}. A consequence of this assumption is the requirement that, in quenched, massless limit, any vertex Ansatz should satisfy the transverse condition~\cite{DMR94,BR93}
\begin{equation}
\int \frac{d^3q}{(2 \pi)^3} D_{\mu \nu}^T \gamma_\mu S(q) \Gamma_\nu (q,p) S(p) = 0.   \label{transversecondition}
\end{equation}
The bare fermion propagator in QED$_3$ transforms under the Landau - Khalatnikov transformation~\cite{LK56} from Landau gauge as follows
\begin{equation}
S(p,0) = \frac{1}{i \not\!p} \Longrightarrow S(p,\xi) = \frac{1}{i \not\! p} \left[1 - \frac{e^2 \xi}{8 \pi p} \arctan \left( \frac{8 \pi p}{e^2 \xi} \right) \right] = \frac{1}{i \not\! p A(p)}.   \label{fermionpropagator}
\end{equation}
The right hand side of (\ref{fermionpropagator}) is the solution of (\ref{fermionSDE2}) provided the transverse condition (\ref{transversecondition}) is satisfied~\cite{DMR94,BR93}.

We write the transverse part of the vertex as 
\begin{equation}
\Gamma_\mu^T (p,q) = \frac{A(p) - A(q)}{p^2 - q^2} \sum_{i = 1}^8 T_\mu^i (p,q) f^i (p^2,q^2,p \cdot q),    \label{transversevertex2}
\end{equation}
where the transverse tensors $T_\mu^i$ are given in the Appendix. Inserting (\ref{fullvertex}) and (\ref{transversevertex2}) into (\ref{transversecondition}), the transverse condition then becomes
\begin{eqnarray}
\int \frac{d^3q}{(2 \pi)^3} & \left\{ D_{\mu \nu}^T (p - q) {\rm tr} \left[ \gamma_\mu S(q) \Gamma_\mu^{BC} (q,p) S(p) \right] \right. \nonumber \\
& + \left. \frac{1}{(p - q)^2} F(q^2, p^2, p \cdot q) \right\} = 0,  \label{transversecondition2}
\end{eqnarray}
where
\begin{eqnarray}
\lefteqn{F (q,p) = {\rm tr} \left[ \gamma_\mu S(q) \Gamma_\mu^T (q,p) S(p) \right]} \nonumber \\
 & = & 4 \frac{A(p) - A(q)}{p^2 q^2 (p^2 - q^2) A(p) A(q)} \left\{ [p^2q^2 - (p \cdot q)^2] \tilde{f} (q,p) \right.  \label{functionF} \\
 & - & \left. 2 [p^2q^2 + (p \cdot q)^2 - (p^2 + q^2) p \cdot q] f^3 (q,p) + 2 p \cdot q (q^2 - p^2) f^6 (q,p) \right\}, \nonumber
\end{eqnarray}
and $\tilde{f} (q,p) = (p^2 + q^2) f^2 (q,p) + f^8 (q,p)$. Assuming the $f^i (q,p)$ are independent of $p \cdot q$~\cite{DMR94} and integrating over angles, we find a simple solution of (\ref{transversecondition2}) 
\begin{equation}
\tilde{f} (q,p) = -2(1 + \beta) \frac{I}{J} \hspace{4mm} ; \hspace{4mm} f^3 (q,p) = - \beta \frac{I}{J} \hspace{4mm} ; \hspace{4mm} f^6 (q,p) = 0,   \label{solution}
\end{equation}
where
\begin{eqnarray}
I (q^2, p^2) & = & \frac{(p^2 + q^2)^2}{16 p q} \ln \left[ \frac{(p + q)^2}{(p - q)^2} \right] - \frac{1}{4} (p^2 + q^2), \nonumber \\
J (q^2, p^2) & = & \frac{(p^2 - q^2)^2}{16 p q} \ln \left[ \frac{(p + q)^2}{(p - q)^2} \right] - \frac{1}{4} (p^2 + q^2). \label{solution2}
\end{eqnarray}
and $\beta$ is an arbitrary parameter. The vertex Ansatz proposed by Dong et al.~\cite{DMR94} is recovered by setting $\beta = 1$.

\section{Photon Polarization Scalar as a Test Case}
\label{Photonscalar}

The validity of the vertex Ansatz (\ref{solution}) can be tested using the photon polarization scalar, which, even in the quenched limit $N_f \rightarrow 0$, is a gauge invariant quantity. The photon polarization tensor in QED$_3$ is
\begin{eqnarray}
\Pi_{\mu \nu} (p) & = & e^2 \int \frac{d^3}{(2 \pi)^3} {\rm tr} \left\{ \gamma_\mu S \left(q + \mbox{$\frac{1}{2}$}p \right) \Gamma_\nu \left(q + \mbox{$\frac{1}{2}$} p, q - \mbox{$\frac{1}{2}$} p \right) \right. \nonumber \\
 & \times & \left. S \left(q - \mbox{$\frac{1}{2}$} p \right) \right\}, \label{photontensor}
\end{eqnarray} 
and the associated photon polarization scalar is
\begin{equation}
\Pi (p) = - \frac{e^2}{2 p^2} \left( \delta_{\mu \nu} - 3 \frac{p_\mu p_\nu}{p^2} \right) \Pi_{\mu \nu} (p). \label{photonscalar}
\end{equation}
Inserting (\ref{fullvertex}) into (\ref{photonscalar}) and using (\ref{functionF}), we get
\begin{eqnarray}
\Pi (p) & = & - \frac{e^2}{2 p^2} \int \frac{d^3q}{(2 \pi)^3} \left\{F \left(q + \mbox{$\frac{1}{2}$} p, q + \mbox{$\frac{1}{2}$} p \right) \right. \nonumber \\
 & + & \left.  {\rm tr} \left[ \gamma_\mu S \left( q + \mbox{$\frac{1}{2}$} p \right) \Gamma_\mu^{BC} \left(q + \mbox{$\frac{1}{2}$} p, q - \mbox{$\frac{1}{2}$} p \right) S \left(q - \mbox{$\frac{1}{2}$} p \right) \right] \right. \nonumber \\
 & + & \left. \frac{3i}{p^2} {\rm tr} \left[ \not\! p \left( S \left( q - \mbox{$\frac{1}{2}$} p \right) - S \left(q + \mbox{$\frac{1}{2}$} p \right) \right) \right] \right\} .   \label{photonscalar2}
\end{eqnarray}

In Landau gauge, if the fermion propagator and vertex reduce to their bare forms, the photon polarization scalar takes the form
\begin{equation}
\Pi (p , \xi = 0) = \frac{e^2}{8 p} .     \label{photonscalar3}
\end{equation}
Since the photon polarization scalar is gauge invariant~\cite{BR93,LK56}, the photon polarization scalar in any arbitrary gauge should take the form of (\ref{photonscalar3}). 

\begin{figure}[t]
\psfig{figure=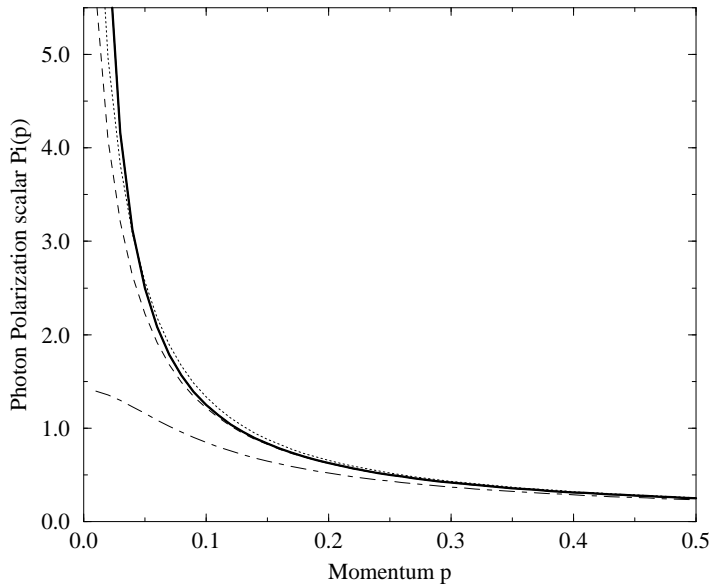,height=90mm}
\caption{The plot of vacuum polarization scalars. The solid line is one loop result (\ref{photonscalar3}). The dash-dotted, dashed and dotted lines are vacuum polarizations in Feynman gauge for $\beta$ = -2/3, 0.5942 and 1 (i.e. the vertex Ansatz proposed by Dong et al.) respectively.     \label{fig1}}
\end{figure}

Inserting (\ref{fermionpropagator}), (\ref{functionF}), (\ref{solution}) and (\ref{solution2}) into (\ref{photonscalar2}), we end up with an integral which can be evaluated numerically. Calculations of the photon polarization scalar in Landau gauge ($\xi = 0$) and Feynman gauge ($\xi = 1$) are shown in Figure \ref{fig1}.
The photon polarization scalar calculated using the vertex Ansatz proposed by Dong et al. ($\beta = 1$) does not match exactly the polarization scalar in Landau gauge given by (\ref{photonscalar3}). In fact, our calculated $\Pi (p)$ has logarithmic behavior as $p \rightarrow 0$ which can be described analytically by

\begin{equation}
\Pi (p) = \frac{1}{\xi} - \frac{2}{\xi} \left( \beta + \frac{2}{3} \right) \ln \left(\frac{4 \pi p}{e^2 \xi} \right) - \frac{1}{\xi} \left( \beta + \frac{2}{9} \right) + O (p \ln p).     \label{photonscalar4}
\end{equation}
Furthermore, $\Pi (p)$ has the following form in ultraviolet regime :
\begin{eqnarray}
\Pi (p) & = &\frac{e^2}{8 p} + \left( - \frac{ \ln 2}{8 \pi^2} + 6.768 \times 10^{-3} + 3.384 \times 10^{-3} \beta \right) \frac{e^4 ( \xi - \xi_0)}{p^2} \nonumber \\
 & + & \left( - \frac{1}{256 \pi^2} + 3.957 \times 10^{-4} \right) \frac{e^6 (\xi - \xi_0)^2}{p^3} + O \left( \frac{1}{p^5} \right).    \label{photonscalar5}
\end{eqnarray}
For $\beta = 0.5942$, (\ref{photonscalar5}) has the same ultraviolet behaviour as (\ref{photonscalar3}) as shown by the dashed line plot in Fig. \ref{fig1}.

\section{Summary and Conclusion}
\label{summary}

We have discussed the use of the photon polarization scalar for testing the gauge covariance of a given fermion-photon vertex Ansatz. As a test case, we have considered a one parameter class of vertex Ans\"{a}tze including that proposed by Dong et al. We find that, in Feynman gauge, the associated polarization scalar has logarithmic behaviour in the infrared regime. By an appropriate parameter choice, we are able to construct an Ansatz which agrees well with the required result (\ref{photonscalar3}) except for small $p$. 

We mention two possible shortcomings of our proposed Ansatz. Firtstly, we note that our class of Ans\"{a}tze has a logarithmic singularity when $k^2 = l^2$, but $k_\mu \not= l_\mu$. It is extremely difficult, however, to construct functions $f^i$ satisfying (\ref{transversecondition2}) which avoid this defect. Secondly, we recall the assumption that the SDEs admit trivial solution in the quenched, massless limit. This assumption has recently been questioned by Bashir et al.~\cite{BKP98}, who claim that the vertex should be augmented by further addition of transverse piece which cannot be determined from properties of single particle propagators alone. 

\section*{Acknowledgement}
PCT would like to acknowledge financial assistance of an Australian Development Co-operation Scholarship from AusAID and an Australian National University Scholarship. The authors would also like to thank the Special Research Centre for Subatomic Structure of Matter in Adelaide where part of this work was completed.

\section*{Appendix}
Of the eight tensors spanning the transverse part of the fermion-photon vertex function, only those consisting of terms containing an odd number of Dirac matrices contribute in the chirally symmetric case. They are given by

\begin{eqnarray}
T_\mu^2 (p,q) & = & (\not\! p + \not\! q)[p_\mu q \cdot(p - q) - q_\mu p \cdot (p - q)]; \nonumber \\
T_\mu^3 (p,q) & = & \gamma_\mu (p - q)^2 - (p - q)_\mu (\not\! p - \not\! q); \nonumber \\
T_\mu^6 (p,q) & = & \gamma_\mu (p^2 - q^2) - (p + q)_\mu (\not\! p - \not\! q); \nonumber \\
T_\mu^8 (p,q) & = & \mbox{$\frac{1}{2}$} [\not\! p \not\! q \gamma_\mu - \gamma_\mu \not\! q \not\! p].    \label{transversefunction}
\end{eqnarray}

\end{document}